\documentclass[aps,twocolumn,showpacs]{revtex4-1}
\usepackage{graphicx}
\usepackage{textcomp}
\usepackage{psfrag}

\usepackage{amssymb}
\usepackage{amsmath}
\usepackage{relsize}
\usepackage{graphicx}% Include figure files
\usepackage[caption=false]{subfig}
\usepackage{dcolumn}% Align table columns on decimal point
\usepackage{bm}% bold math

\begin{document}
%\openup 1em

\title{Energy Efficiency of Hydrocarbon Isomer Separation via Levi–Blow Mechanism: Benchmarking Against Conventional Methods}

\author{Shubhadeep Nag} 
\email{shubhadeepn@iisc.ac.in/shubhadeepnag92@gmail.com}
\author{Yashonath Subramanian}
\affiliation{Solid State and Structural Chemistry Unit, Indian Institute of Science, Bangalore - 560012}

\begin{abstract}
Separation of hydrocarbon isomers with closely matched physicochemical properties remains an energetically demanding challenge in the petrochemical refineries. This study presents a comparative analysis of three methods: the fractional distillation, the Molex process, and the Levi–Blow (LB) technique.
Using experimentally validated thermodynamic data and heat-transfer estimates, we quantify the energy expenditure and number of cycles required to achieve ultrahigh-purity separation. The LB method achieves eight-9s purity in a single cycle with just 39.4~kJ/mol, while fractional distillation and Molex require multiple iterations and up to five orders of magnitude more energy.
Each method is also benchmarked against the thermodynamic minimum work of separation of the mixture of interest. It is found that the LB technique operates within a factor of 23 of this lower bound, compared to $\sim 10^2$ for Molex \footnote{Molex is a trademark and/or service mark of UOP Inc.} and $\sim 10^5$ for fractional distillation, demonstrating its thermodynamic and operational superiority.
%Our findings demonstrate that the Levi--Blow technique represents a transformative advance in separation science, offering ultrahigh purity with drastically reduced energy cost. This positions the method as a highly promising, green alternative for next-generation hydrocarbon separation technologies.
\end{abstract}

\date{\today}% It is always \today, today,

\maketitle

\section{Introduction}
The green-chemistry separation of molecular mixtures is key to overcoming the excessive use of 
fossil fuels and controlling the emission of greenhouse gases, but at present, no such industrial-scale green 
chemistry separation technique can replace the existing energy-intensive methods \cite{Afonso2005Green, Shao2020_c1, Hussain2025, Shang2025}. Among industrial operations, the petroleum refining and chemical separation industries represent some of the most energy-intensive sectors, with separation techniques such as fractional distillation alone estimated to consume nearly 15$\%$ of the world’s total energy \cite{Kiss2020, Sholl2016, Cho2024}. This vast energy footprint stems from the repeated heating and phase-change operations required to isolate components from complex molecular mixtures, particularly hydrocarbon streams \cite{osti_c1}. In light of ongoing climate change and increasingly ambitious decarbonization targets, there is mounting pressure to rethink how separations are performed at an industrial scale \cite{Zhao2024}. While shifts toward renewable energy sources are essential, these transitions are gradual and insufficient to meet global emission reduction targets in the near term \cite{yang2024}. As such, developing radically more efficient separation methods presents a critical and timely opportunity for scientific and technological intervention.

Conventional separation techniques, including distillation, adsorption-based processes, membrane filtration, and solvent extraction, operate based on differences in equilibrium thermodynamic or kinetic properties of the components in a mixture \cite{seader, Ruthven1984}. For instance, in fractional distillation, differences in boiling point and vapor pressure are exploited: the more volatile component evaporates preferentially and is subsequently condensed \cite{Donahue2002}. However, this approach has two major limitations. First, the external parameters applied (e.g., heat, pressure) tend to shift the properties of all components in the same direction, which inherently limits the maximum achievable separation factor. Second, achieving high-purity outputs typically requires multiple stages or cycles, each incurring additional energy costs \cite{cen, Ronald}. These issues become even more acute when the components are structurally or chemically similar, as in the case of neopentane and $n$-pentane, or $n$-hexane and 2,2-dimethylbenzene, or similar branched-linear isomers. In such cases, even advanced techniques like cryogenic distillation or membrane-based separations struggle to achieve high selectivity without prohibitive energy consumption \cite{Porter2000}.

Given these constraints, there is a compelling need for a separation mechanism that transcends the standard frameworks of selective adsorption or vapor–liquid equilibrium. Such a mechanism should be capable of introducing a directional bias into the mixture, selectively activating distinct molecular properties in ways that amplify their differences. Unlike equilibrium-based approaches, this strategy would actively discriminate between molecular species, thereby facilitating their segregation with greater efficiency.
For the traditional diffusion based approach in zeolites and other MOFs, it is a practice to separate molecules based on their relative diffusivities of different molecular species in opposite directions, rather than relying on passive partitioning or bulk property gradients \cite{Smit2008, Chmelik2010, Krishna2012}. Recent studies have revealed that diffusion in zeolites is far more complex than simple size-exclusion or Fickian scaling, with mechanisms such as molecular self-gating controlling transport under confinement \cite{Liu2025}. Counterintuitive coupling between translation and rotation under confinement has also been demonstrated, where smaller asymmetric molecules diffuse more slowly than larger symmetric ones due to selective pore–guest interactions \cite{Liu2025a}. Furthermore, confinement can invert classical expectations of temperature–diffusion scaling, as shown by the recently identified thermal resistance effect, where diffusion of long-chain hydrocarbons slows with increasing temperature \cite{Yuan2021}. Structural motifs such as continuum intersecting channels in zeolites have also been shown to synergistically enhance diffusion pathways, overcoming the adsorption–diffusion trade-off and enabling selective transport \cite{Liu2021}. Such advances underscore that biased, non-equilibrium separation strategies in porous frameworks are both physically realizable and technologically promising.

In this context, we focus on a recently proposed approach known as the Levi–Blow (LB) separation technique, which combines two counterintuitive physical effects to achieve efficient, directional separation of molecular mixtures \cite{Nag2020, Nag2021}. The first is the Levitation Effect (LE), wherein molecular dynamics simulations and theoretical studies have shown that the diffusivity of a guest molecule in a porous medium exhibits a non-monotonic dependence on its size \cite{Santi1994, sanjoy1995}. Specifically, as the size of the guest approaches the bottleneck diameter of the pore (e.g., the window of a zeolite), its diffusivity peaks sharply \cite{Yasho2008}. This is attributed to the mutual cancellation of host–guest forces when the guest is symmetrically situated within the pore throat, leading to lower activation energy barriers for diffusion \cite{Ghorai2005}. The second ingredient is the Blowtorch Effect (BT), introduced by Landauer in his seminal work in 1975 \cite{Landauer1975}, which involves creating a localized hot zone in a bistable or periodic potential \cite{Landauer1988}. This localized heating effectively flattens the energy barriers in that region, allowing a molecule to overcome them more easily and thereby creating biased diffusion \cite{Kampen1988, Ray2015, Ray2015a}. When these two effects are judiciously combined, e.g., placing a hot zone near the pore window in a crystalline host, it becomes possible to drive different components of a mixture in opposite directions, resulting in ultra-high selectivity and minimal energy expenditure \cite{Nag2021a}.

While the conceptual mechanism of the LB technique has been validated in prior computational studies, including successful separation of neopentane/$n$-hexane and $n$-pentane/2,2-dimethylbutane mixtures — the question of practical implementation and scalability remains largely unexplored \cite{Nag2022, Nag2022a}. In particular, there has been no prior work that quantifies the energy cost of achieving a given separation factor via the LB method, nor any effort to benchmark this against existing industrial techniques. To address this gap, the present work performs a comprehensive energy and separation factor analysis of the LB technique, using both thermodynamic data and simulation-driven estimates of diffusion and selectivity. We compare the LB approach directly to fractional distillation and the Molex process, two widely used techniques for alkane isomer separation, focusing on neopentane/$n$-pentane as a representative test case.

%Specifically, we compute the energy expenditure per mole of mixture for each method, accounting for the number of iteration cycles required to reach a target purity (e.g., 99.9$\%$), as well as the associated separation factor per cycle. Our results show that the LB method achieves orders of magnitude higher separation factors with as little as a 40 K temperature gradient, and with energy inputs in the range of 0.1–0.5 kJ/mol, in stark contrast to the 20–50 kJ/mol typically required by distillation. These findings not only validate the LB method as a highly energy-efficient alternative, but also provide critical insight into its feasibility for industrial-scale implementation, laying the groundwork for experimental realization and techno-economic analysis.

\section{Levi-Blow Effect}
\subsection{Levitation Effect}

Molecular diffusion in porous crystalline materials such as zeolites, BCC, FCC,  CNTs, etc., exhibits a strikingly non-monotonic dependence on the size of the diffusing guest molecule \cite{Krishna2009, Subramanian2020-06-18}. This phenomenon, known as the \textit{Levitation Effect}, has been observed across a range of systems, including liquids, amorphous solids, and close-packed crystals, and is particularly relevant in the context of diffusion within confined porous frameworks \cite{Ghorai2003, Smit2006, Nag2020JCP}. Derouane and co-workers provided a theoretical explanation for this behavior by analyzing curvature-induced force cancellation within the pore \cite{Derouane1987, Derouane1988}.

\begin{figure}
    \centering
    \includegraphics[width=0.95\linewidth]{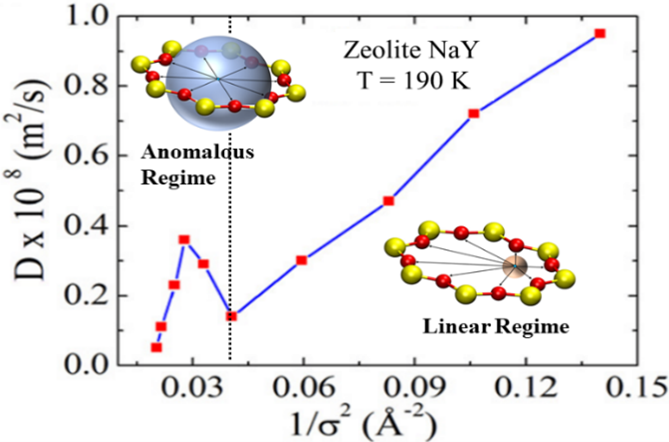}
    \caption{Diffusivity as a function of $1/\sigma^{2}$ in zeolite NaY \cite{Santi1994}, where $\sigma$ represents the diameter of the guest molecule.}
    \label{Levi}
\end{figure}

It was demonstrated that the self-diffusivity of a guest molecule varies anomalously as a function of its van der Waals diameter $\sigma_{gg}$ when transported through a microporous host (see Fig.~\ref{Levi}). For small guest sizes, diffusivity follows the expected inverse-square trend given by the Stokes–Einstein relation:
\begin{equation}
D = \frac{RT}{N_A \cdot 6\pi \eta r_g},
\end{equation}
where $D$ is the diffusivity, $R$ is the universal gas constant, $T$ is temperature, $N_A$ is Avogadro’s number, $\eta$ is the viscosity, and $r_g$ is the radius of the guest molecule. This regime, where diffusivity decreases with increasing guest size, is termed the linear regime (LR).

However, as the guest diameter approaches the characteristic bottleneck diameter $\sigma_w$ of the pore network, such as the window between two cages in a zeolite, the diffusivity unexpectedly increases and reaches a pronounced maximum. This counterintuitive increase defines the anomalous regime (AR). Physically, this effect is attributed to the symmetry-induced cancellation of forces experienced by the guest from the host atoms when the guest is centrally located within the pore throat. In this configuration, the net mean force acting on the molecule is minimized, resulting in a flatter potential energy landscape and, consequently, a lower activation barrier for translational motion.

% When the guest size matches the bottleneck, the repulsive and attractive van der Waals interactions from opposing walls nearly cancel, leading to reduced energy undulations along the diffusion path.

To quantify this phenomenon, the \textit{levitation parameter}, $\gamma$ is introduced as:
\begin{equation}
\gamma = \frac{2 \cdot 2^{1/6} \cdot \sigma_{uv}}{\sigma_w},
\end{equation}
where $\sigma_{uv}$ represents the effective solute–solvent (guest–host) interaction diameter, and $\sigma_w$ is the diameter of the pore bottleneck. The levitation effect is observed when $\gamma \approx 1$, corresponding to the case where the guest diameter is optimally matched to the pore window.

This diffusivity peak is not only a universal consequence of host–guest interaction geometry but also a highly desirable feature for separation applications. By selecting a host whose pore size matches one of the components in a mixture, it becomes possible to exploit the levitation effect to selectively enhance the mobility of that component while suppressing that of others. In 2010, Borah and co-workers experimentally demonstrated the Levitation Effect by measuring the diffusivity of neopentane, isopentane, and $n$-pentane in zeolite Y using QENS. They observed that neopentane exhibits the highest diffusivity, followed by isopentane, and then $n$-pentane. \cite{Borah2010}. %When combined with an external driving field, such as a localized hot zone, this differential mobility serves as the foundation for directional, energy-efficient separations using the Levi–Blow approach.

\subsection{Blowtorch Effect}

The \textit{Blowtorch Effect}, first introduced by Landauer, describes a counterintuitive thermodynamic phenomenon wherein the equilibrium distribution of particles in a bistable potential can be significantly altered by applying localized heating in a specific region of the energy landscape \cite{Landauer1975}. This effect challenges the conventional expectation that particles preferentially occupy lower-energy states under thermal equilibrium.

\begin{figure}
    \centering
    \includegraphics[width=0.95\linewidth]{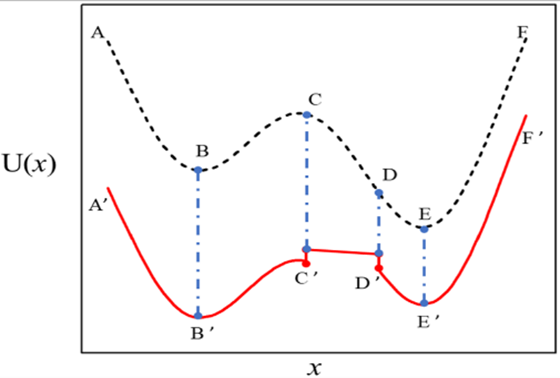}
    \caption{Schematic representation of the blow-torch effect.}
    \label{BT}
\end{figure}

Consider a bistable potential energy landscape with two local minima, labeled $E$ and $B$, separated by an energy barrier with a maximum at point $C$ (see Fig.~\ref{BT}). Under uniform thermal conditions at ambient temperature $T_a$, the equilibrium populations in each well follow the Boltzmann distribution:
\begin{equation}
p(x) \propto \exp\left(-\frac{U(x)}{k_B T_a}\right),
\end{equation}
where $U(x)$ is the potential energy at position $x$, and $k_B$ is the Boltzmann constant. If $E$ is lower in energy than $B$, the system will naturally favor population of the $E$ well.

However, Landauer demonstrated that introducing a \textit{hot zone}—a spatial region with elevated temperature $T_h > T_a$—between points $C$ and $E$ (i.e., in the uphill region of the lower well), can dramatically modify this distribution. In the hot zone, the effective potential energy can be redefined using the inverse Boltzmann relation:
\begin{equation}
U(x) = -k_B T_h \ln p(x).
\end{equation}
As $T_h > T_a$, the energy barrier in the heated region becomes effectively \textit{flattened}, leading to a reduced activation barrier between $E$ and $B$. This causes an enhanced transition rate toward the higher-energy state $B$, resulting in an anomalous population inversion where $p(x_B) > p(x_E)$ despite $U(B) > U(E)$.

This phenomenon, termed the \textit{blowtorch effect}, has since been investigated in depth using both analytical and simulation-based approaches. van Kampen and B\"{u}ttiker reformulated Landauer's insight within a rigorous statistical thermodynamic framework \cite{Kampen1988, Buttiker1998}. Subsequent studies by Bekele and Ray further elucidated the dynamical consequences of applying non-uniform temperature fields to bistable and periodic potentials. For instance, it was shown that varying the position and width of the hot zone leads to complex changes in the stationary distribution, including the appearance of double maxima and nonmonotonic trends in population shifts.

The blowtorch effect is particularly significant in systems where controlled directional transport is desired without relying on mechanical or external field gradients. In the context of molecular separation, placing a hot zone at a strategic location within a periodic potential or a porous framework creates a directional bias in diffusion. This breaks the symmetry of the system and enables selective enhancement of transport for certain molecular species over others.

%When combined with the levitation effect—which already provides species-dependent mobility due to size-matching with the pore—the blowtorch effect serves as a powerful non-equilibrium driving mechanism. The synergy of these two phenomena underpins the Levi–Blow separation method, enabling molecular components in a mixture to be \textit{actively driven} in opposite directions with minimal energy input, thus achieving exceptional separation performance.

\subsection{Levi--Blow Method}

Having introduced both the levitation and blowtorch effects, we now describe how these two phenomena can be judiciously combined to achieve directional separation of molecular mixtures. We illustrate the approach using an equimolar mixture of neopentane and \emph{n}-pentane confined in the zeolite NaY, which serves as the host material. The detailed results are discussed in Nag \emph{et al.} \cite{Nag2021}. The system is initialized with one molecule of each component per cage, and simulations are carried out to observe how each species responds to the applied thermal and spatial modulation.

\begin{figure}[h]
    \centering
    \includegraphics[width=0.45\textwidth]{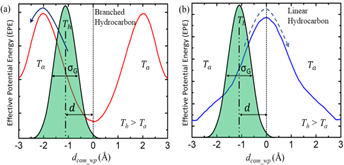}
    \caption{Schematic representation of the Levi--Blow technique. The hot zone and temperature $T_h$ displaces anomalous-regime molecules leftwards and linear-regime molecules rightwards.}
    \label{fig:epelandscape}
\end{figure}

The window diameter of zeolite NaY is approximately 7.4~\AA. The kinetic diameter of neopentane is 6.2~\AA, which places it in the anomalous or levitating regime. In contrast, the kinetic diameter of \emph{n}-pentane is 4.3~\AA, situating it in the linear regime of the levitation profile. As a result, neopentane experiences a potential minimum at the window due to force cancellation effects, while \emph{n}-pentane experiences a potential maximum. The schematic potential energy profiles experienced by neopentane (in red) and \emph{n}-pentane (in green) are shown in Fig.~\ref{fig:epelandscape}.

\begin{figure}[h]
    \centering
    \includegraphics[width=0.45\textwidth]{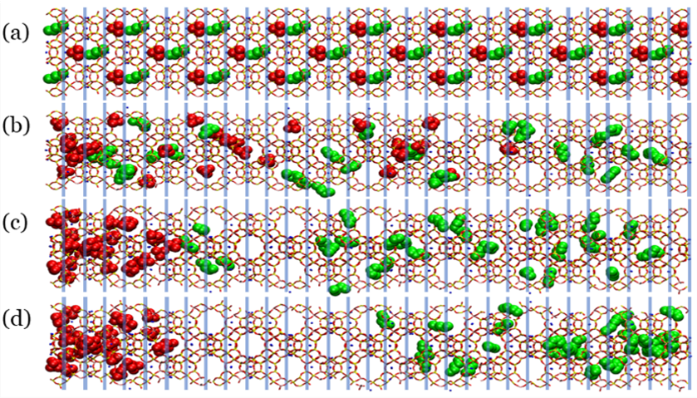} % Replace with correct image filename
    \caption{Snapshots of the separation of the equimolar mixture of neopentane and \emph{n}-pentane in an $8 \times 1 \times 2$ zeolite NaY at (a) initial, (b) 0.1 million, (c) 0.5 million, and (d) 3.1 million MC steps. The vertical shaded regions indicate the Gaussian hot zone \cite{Nag2021}.}
    \label{fig:sepsnaps}
\end{figure}

To activate directional separation, we impose a Gaussian-shaped hot zone slightly away from the midpoint between the two potential wells. The elevated temperature in this region modifies the effective potential energy landscape via the blowtorch effect, reducing the barrier between minima and biasing the net molecular flux. Specifically, \emph{n}-pentane, being in the linear regime, is pushed toward the right, while neopentane in the levitating regime is driven leftward. This results in the two species migrating in opposite directions away from each other.
Snapshots of the system at different Monte Carlo steps are shown in Fig.~\ref{fig:sepsnaps}. The separation begins with a uniform initial distribution and evolves toward complete segregation at the end of the simulation under the influence of the combined levitation and blowtorch mechanisms. Notably, this separation is achieved within a single simulation cycle, highlighting the efficiency and directional nature of the Levi-Blow technique.

These results demonstrate that a carefully engineered hot zone, when applied to a host–guest system with size-selective diffusivity behavior, can achieve complete separation of molecular mixtures using directional thermally biased transport. The Levi--Blow method thus offers a novel, low-energy, and high-selectivity alternative to traditional separation processes.

%% Section 4: Cost Analysis and Iterative Energy Comparison for Pentane Isomer Separation

\section{Energy Cost and Iterative Cycles of Different Methods}

Recent work highlights the critical role of techno–economic analysis (TEA) in assessing the viability of emerging separation and conversion technologies. For adsorption-based processes, TEA models derived from breakthrough experiments enable accurate prediction of isotherms and process economics \cite{Thijs2019}. In the context of CO$_2$ electroreduction, multiple studies have shown that issues such as carbonate crossover, electrolyte recovery, and low single-pass utilization strongly dictate commercial feasibility \cite{Sisler2021, Ramdin2023}. Comparative evaluations of direct versus two-step CO$_2$–to–C$_2$ pathways further emphasize that profitability hinges on cell voltage, capital cost, and electricity price \cite{Ramdin2021}. Collectively, these examples emphasize that rigorous TEA is required not only for novel processes but also for benchmarking emerging approaches against incumbent industrial methods. This motivates our quantitative analysis of energy expenditure and cycle requirements for Levi–Blow relative to distillation and Molex.

\begin{table}[h]
\caption{
Thermodynamic Properties of Pentane Isomers\cite{Good1970}}
\begin{ruledtabular}
\begin{tabular}{cccccc}
Molecule&Density&$\Delta$H$_{vap}$&T$_{\text{B.P.}}$&C$_{p}$&P$_{vap}$ \\
&g cm$^{-3}$&kJ mol$^{-1}$&$^{o}$C&J K$^{-1}$ mol$^{-1}$&kPa(20$^{o}$ C) \\
\hline
neopentane& 0.621& 26.66& 9& 120.00&57.9 \\
isopentane& 0.621& 24.94& 28& 164.85&79.0 \\
$n$-pentane& 0.621& 21.93& 36& 167.19&146.0 \\
\end{tabular}
\end{ruledtabular}
\label{prop_table}
\end{table}

\begin{table*}
\caption{\label{tab:table2}
Calculated mole fractions, separation factors, and cumulative energy consumption for neopentane/$n$-pentane separation by fractional distillation as a function of iteration cycle.}
\begin{ruledtabular}
\begin{tabular}{cccccc}
\textbf{Iteration}&$\it{\textbf{x}}$&$\it{y}$&$\alpha$&$\beta$&E (kJ/mol) \\
\hline
0(initial)&0.5&0.5&0.5&0.5&0.0\\
1&0.355&0.145&0.71&0.29&12.57\\
8(one-9)&3.33$\times$10$^{-2}$&2.5$\times$10$^{-5}$&0.999&7.74$\times$10$^{-4}$&34.77\\
11(two-9)&1.15$\times$10$^{-2}$&6.1$\times$10$^{-5}$&0.9999&5.28$\times$10$^{-5}$&36.18\\
16(four-9)&2.08$\times$10$^{-3}$&1.25$\times$10$^{-5}$&0.9999&5.36$\times$10$^{-7}$&36.43\\
21(six-9)&3.76$\times$10$^{-4}$&2.56$\times$10$^{-5}$&0.999999&6.81$\times$10$^{-9}$&36.84\\
28(eight-9)&3.42$\times$10$^{-5}$&4.46$\times$10$^{-5}$&0.99999999&1.29$\times$10$^{-11}$&36.86\\
\end{tabular}
\end{ruledtabular}
\end{table*}

Separating structurally similar hydrocarbon isomers, such as neopentane and \textit{n}-pentane, poses a significant industrial challenge due to their closely matched physical properties. Traditional separation techniques like fractional distillation and the Molex process require considerable energy and often involve multiple iterative cycles to achieve high-purity separation. In this section, we present a quantitative energy comparison across these conventional methods, benchmarked against the Levi--Blow (LB) technique proposed in this study.

Here, in this section, we are going to calculate the energy cost of different methods to separate neopentane and $n$-pentane mixtures. There are different methods to separate hydrocarbon mixtures. Fractional distillation and Molex process are among of them. We have calculated here the energy cost of these methods in detail and compared it with our proposed Levi-Blow method. The thermodynamic properties of pentane isomers are listed in Table~\ref{prop_table}.

\begin{table*}
\caption{\label{tab:table3}
Mole fraction profiles and corresponding energy expenditure across iterative cycles of Molex for neopentane/$n$-pentane.}
\begin{ruledtabular}
\begin{tabular}{cccccc}
Iteration&$\it{x}$&$\it{y}$&$\alpha$&$\beta$&E(kJ/mol) \\
\hline
0(initial)&0.5&0.5&0.5&0.5&0.0\\
1&0.49&	9.99$\times$10$^{-3}$&	0.98&	0.02&	51.87\\
2(one-9)&	0.48&	1.99$\times$10$^{-4}$&	0.999&	4.16$\times$10$^{-4}$&	103.74\\
3(three-9)&	0.47&	3.99$\times$10$^{-6}$&	0.99999&	8.49$\times$10$^{-6}$&	155.61\\
4(four-9)&	0.46&	7.99$\times$10$^{-8}$&	0.999999&	1.73$\times$10$^{-7}$&	207.48\\
5(six-9)&	0.45&	1.59$\times$10$^{-9}$&	0.99999999&	3.54$\times$10$^{-9}$&	259.35\\
6(eight-9)&	0.44&	3.19$\times$10$^{-11}$&	0.9999999999&	7.22$\times$10$^{-11}$&	311.22\\
\end{tabular}
\end{ruledtabular}
\end{table*}

\subsection{Fractional Distillation}

Fractional distillation separates components of a liquid mixture based on their relative volatilities, typically characterized by differences in vapor pressure or boiling points. The process involves continuously vaporizing the mixture and passing the vapor through a vertical distillation column packed with trays or structured internals that facilitate repeated vapor–liquid equilibrium stages. Lighter components with higher vapor pressures rise and condense at higher points in the column, while heavier, less volatile components condense lower down. 

For a binary mixture like neopentane (\(T_b = 9^\circ\text{C}, P_{\text{vap}} \approx 1.0~\text{atm}\)) and \textit{n}-pentane (\(T_b = 36^\circ\text{C}, P_{\text{vap}} \approx 0.4~\text{atm}\)), the small boiling point difference (\(\Delta T_b \approx 27^\circ\text{C}\)) results in relatively low relative volatility (\( \alpha < 2.5 \)), necessitating a large number of theoretical stages to achieve sharp separation. Each vaporization-condensation cycle consumes latent heat, and to reach purities beyond 99.9999\%, hundreds to thousands of cycles may be required. 

In large-scale applications such as petroleum refining, distillation columns may span 30–60 meters in height and operate at pressures ranging from 1 to 30 atm, depending on the feed composition and cut points. The energy consumption is substantial, with distillation accounting for nearly 15\% of industrial energy usage globally. In our case study, achieving 8-nine (99.999999\%) purity of neopentane from an equimolar mixture required over 409,000 iteration cycles, translating to more than 0.5 MJ/mol energy usage, underscoring the inefficiency of distillation for closely related isomers.

We assume that initially neopentane and $n$-pentane are present in a 50:50 composition in the mixture. The vapor pressure of neopentane is $\thicksim$ 1.0 atm and $n$-pentane is $\thicksim$ 0.4 atm at 300K.
In 300 K, neopentane is in gaseous state (Boiling point of neopentane = 9$^{o}$C). Therefore, the composition of evaporated(distilled) mixture is

\begin{equation}
\begin{split}
n-\text{pentane} \simeq \frac{0.4}{1+0.4};   
\text{neopentane} \simeq  \frac{1}{1+0.4}
\end{split}
\end{equation}

Considering ideal gas behaviour, the energy expenditure for distillation is
\begin{equation}
\begin{split}
\simeq 0.5\times27\text{kJ/mol}+0.5\times22\text{kJ/mol}\\
= 24.5\text{kJ/mol}
\end{split}
\end{equation}

\begin{table*}
\centering
\caption{Comparison of Actual Energy Cost vs Minimum Thermodynamic Work}
\label{tab:minwork}
\begin{tabular}{lccc}
\hline
\textbf{Method} & \textbf{Actual Energy} (kJ/mol) & \textbf{W$_{\text{min}}$ (kJ/mol)} & \textbf{Efficiency (\%)} \\
\hline
Fractional Distillation & 538888.89 & 1.73 & 0.00032 \\
Molex Process & 353.66 & 1.73 & 0.49 \\
Levi--Blow Method & 39.42 & 1.73 & 4.39 \\
\hline
\end{tabular}
\end{table*}

Suppose, neopentane and  $n$-pentane composition are $\it{x}$ and $\it{y}$ mole respectively.
Relative proportion of neopentane and $n$-pentane in the mixture is 
\begin{equation}
\alpha=\it{\frac{x}{x+y}}; \beta=\it{\frac{y}{x+y}}
\end{equation}

and energy usage, E in kJ/mol. 
In Table \ref{tab:table2}, the energy consumption of fractional distillation is listed corresponding to the iteration cycle.

After 28th iteration cycle, from 0.5 mole of each neopentane and $n$-pentane mixture, only 3.42$\times$10$^{-5}$ mole of neopentane and 4.46$\times$10$^{-16}$ mole of $n$-pentane yielded with eight 9 (99.99999999 \%) purity in neopentane.
Therefore, percentage of neopentane in 28th cycle = 3.42$\times$10$^{-5}$$\times$100/0.5 $\simeq$ 0.00684 \% of 0.5 mole of initial composition.
Thus, to produce 1 mole of neopentane with eight 9 separation factor, energy is required = $\frac{36.86}{0.0000684}$ $\simeq$538.83 $\times$10$^ {3}$ kJ/mol, and an iteration cycle is needed = $\frac{28}{0.0000684}$ $\simeq$409356.

\subsection{Molex Process}

The Molex process is an industrial adsorptive separation technique developed by UOP for the extraction of linear paraffins from isomeric and cyclic hydrocarbons. It employs fixed-bed columns packed with crystalline zeolites such as 5A or NaY as selective adsorbents. These zeolites possess uniform pore structures (e.g., ~5.1 \AA) that can accommodate linear alkanes like \textit{n}-pentane (kinetic diameter ~4.3 \AA) while excluding branched alkanes such as neopentane (kinetic diameter ~6.2 \AA) due to steric hindrance.

The process operates under elevated temperatures (\(T \approx 170^\circ\text{C}\)) and moderate pressures (4–5 atm), promoting adsorption of linear hydrocarbons onto the zeolite framework. The feed mixture is introduced into the column, where linear molecules are retained and branched molecules are collected in the raffinate stream. Periodically, the column is regenerated via pressure swing or thermal swing desorption, often aided by purge gases or temperature ramps.

Thermodynamically, the energy cost includes heating the hydrocarbon mixture and the entire zeolite column to operational temperature, as well as overcoming adsorption enthalpies (typically 10–30 kJ/mol depending on the zeolite–guest pair). In our benchmark, we estimated energy requirements for a single Molex cycle to be ~71.87 kJ/mol. While the method achieves reasonable selectivity (~98:2 in one pass), multiple cycles are required to reach extreme purities. For 8-nine purity, 6 cycles were necessary, with a cumulative energy cost of ~353.66 kJ/mol. Additionally, the thermal inertia of bulk zeolite structures introduces inefficiencies not present in more localized heating approaches.

In the Molex process, the mixture is heated at 170 $^{o}$C and at 4-5 atm pressure.
Starting at 300K, neopentane is already in the gaseous state.
Therefore, to increase its temperature to 443K (170 $^{o}$C) heat energy required for neopentane is

\begin{equation}
\begin{split} 
=(C_{p}^{neopentane} \times \Delta T)_{gas}\\
=120 JK^{-1} \times (273+170-300) K\\
=17.16kJ mol^{-1}
\end{split}
\end{equation}

At 300K, neopentane is in a liquid state
\begin{equation}
\begin{split}
Q_{n-pentane}=\big(C_{p}^{\textit{n}-pentane} \Delta T \big)_{liq} + \Delta H_{vap} +\\
 \big(C_{p}^{neopentane} \Delta T \big)_{gas}\\
= 167.19 \times (273+36-300) + \\
26\times10^{3} + 167.19 \times (273+170-309) \text{\text{J}}\\
 = 49.91 \text{kJ/mol}
\end{split}
\end{equation}

Therefore, total energy required in one cycle = 19.56 + 52.31 = 71.87 kJ/mol.
In one cycle of the Molex process, 98 \% neopentane and 2 \% $n$-pentane are yielded.
In Table \ref{tab:table3}, energy consumption of the Molex process is listed corresponding to the iteration cycle.
After 6th iteration cycle, from 0.5 mole of each neopentane and $n$-pentane mixture, only 0.44 mole of neopentane and 3.19$\times$10$^{-11}$ mole of $n$-pentane were yielded with eight 9 (99.99999999 \%) purity in neopentane.
Therefore, the percentage of neopentane in the 28th cycle = 0.44$\times$100/0.5=88 $\%$ of 0.5 mole of initial composition.
Thus, to produce 1 mole of neopentane with eight 9 separation factor, energy is requires $\simeq$ 311.22/0.88 $\simeq$ 353.66 kJ/mol, and the number of iteration cycles needed = 6 / 0.88 $\simeq$7.

In the Molex process, hydrocarbons diffuse through the zeolite column. Therefore, we also need to include the energy required to heat the zeolite column. To separate the hydrocarbons, the zeolite column also needed to be heated at 170 $^{o}$ C. 
Consider, the dimension of the zeolite column is 33$\times$1$\times$1$\times$ m$^{3}$.
Therefore, the volume of this zeolite column = $\pi$ $\times$ r$^{2}$ $\times$ h.
The volume of one unit cell of zeolite = (25.8536 m)$^{3}$.
Thus, the number of unit cells in the column = 1.689 $\times$ 10$^{27}$.

Mass of the zeolite column = mass of one unit cell $\times$ total no of unit cells
\begin{equation}
\begin{split}
			   = 12468 \times 1.6 \times 2.112 \times 10^{3}\\
			   = 43.711 \times 10 ^{6} gram
\end{split}
\end{equation}

Therefore, the heat required to maintain the zeolite column at 170$^{o}$C

\begin{equation}
\begin{split}
\Delta Q = 43.711 \times 10^{6} \times 0.8 \times (273+170-300) \text{J} \\
	 = 5.00 \times 10^{9} \text{J}
\end{split}
\end{equation}

\subsection{Levi - Blow Method}

In the Levi-Blow method, we periodically put local hot-zones in zeolite. We maintain the hot temperature at 330K while ambient at 300K.
Energy required for heating neopentane and $n$-pentane respectively

\begin{equation}
\begin{split}
Q_{neopentane}=(C_{p}^{neopentane}\times\Delta T)_{gas}\\
=120 J K^{-1} mol^{-1}\times\big( 330 - 300 \big)K\\
=3.6 kJ mol^{-1}
\end{split}
\end{equation}

%Energy required for $n$-pentane 
\begin{equation}
\begin{split}
Q_{n-pentane}=\big(C_{p}^{\textit{n}-pentane} \Delta T \big)_{liq} + \Delta H_{vap} +\\
 \big(C_{p}^{neopentane} \Delta T \big)_{gas}\\
= 167.19 J K^{-1}mol^{-1}\times\big( 273 + 36 - 300 \big) K + \\
26,000 J + 167.19 J K^{-1}mol^{-1}\times\big( 330 - 309 \big) K\\
= 31.02 kJ mol^{-1}
\end{split}
\end{equation}

Total Energy required to heat hydrocarbons 
\begin{equation}
= 3.6 + 31.02 = 34.62 kJ mol^{-1}
\end{equation}

Consider a zeolite column of 1 $\mu$m $\times$ 0.5 $\mu$m $\times$ 0.5$\mu$m. No of Unit Cells in this column = $\frac{10000}{25}$ $\times$ $\frac{5000}{25}$ $\times$ $\frac{5000}{25}$ =  1.6 $\times$ 10$^{7}$

Number of atoms in one unit cell of zeolite Y = 624
Approximate number of atoms in one hot zone of width 1 \AA in \textit{x}-direction = 624/25 = 25
In zeolite, for one Si, two O atoms are there.
Approximate mass of one hot zone
 
\begin{equation}
\begin{split}
= 8\times m_{Si}+16\times m_{O}\\
= 8 \times 28 amu + 16 \times 16 amu\\
= 480 amu\\
\simeq 7.68 \times 10 ^{-22} gm
\end{split}
\end{equation}

mass of four hot-zone = 3.07 $\times$ 10 $^{-21}$ gram.
Each unit cell consists of 4 hot-zones.
Mass of all hot-zones in  1.6 $\times$ 10$^{7}$ unit cells = 4.9 $\times$ 10 $^{-13}$ gram. The specific heat of zeolite is 0.8 J/(m.K) \cite{White2006}.

Energy to raise the temperature to 330K from 300K
\begin{equation}
\begin{split}
= m \times c_{p} \times \Delta T\\
= 5 \times 10^{-13} \times 0.8 \times (330 - 300) \text{J}\\
= 1.2 \times 10^{-11} \text{J}
\end{split}
\end{equation}

We also need to compute the heat loss from the hot-zones. Below, we have calculated the time required for the hot zone go down at 320 K from 330K. This would tell us how frequently we need to heat zeolite hot-zones to achieve the best separation factor.

Thermal conductivity of zeolite NaY $\simeq$ 0.15 W / (m.K)
heat flux = 0.15 W/m$^{2}$
Therefore, when $\Delta$T/$\Delta$\textit{x} = 30K, heat flux = 4.5 W/m$^{2}$ = 4.5 $\times$10$^{-20}$ \text{J}s/(\AA$^{2}$.sec)
Therefore, 4.5 $\times$10$^{-20}$ \text{J}s heat is released in 1 sec when $\Delta$T/$\Delta$\textit{x} = 30K
Energy required to heat one hot-zone at 30K higher = 1.2 $\times$ 10$^{-11}$ / 1.6 $\times$ 10$^{7}$ \text{J} = 7.5 $\times$ 10 ${-19}$ \text{J}.
To get down of temperature from 330K to 320K , energy radiated = 1/3 $\times$ 7.5 $\times$ 10$^{-19}$ \text{J}
= 2.5 $\times$ 10$^{-19}$ \text{J}.

Now, in 1 sec, through 1 \AA$^{2}$ cross-sectional area, heat flux value = 4.5 $\times$ 10$^{-20}$ \text{J}. The cross-section of zeolite column is 1 $\mu$m $\times$ 0.5$\mu$m $\times$ 0.5$\mu$m is 0.25 $\mu^{2}$. In 1 sec, through 0.25 $\mu^{2}$ cross-sectional area heat flux

\begin{equation}
\begin{split}
= 4.5 \times 10^{-20} \times 0.25 \times 10^{8} \text{J}
\\
\simeq 1.125 \times 10^{-12} \text{J}
\end{split}
\end{equation}

Therefore, 1.125 $\times$ 10$^{-12}$ \text{J} heat radiates in 1 sec.
Now, to radiate 10K temperature by one hot-zone, 2.5 $\times$ 10$^{-19}$ heat loses, and it radiates in 0.22 $\mu$sec.
And, the diffusion coefficient of neopentane and $n$-pentane are 3.17 $\times$ 10 $^{-9}$ m$^{2}$/sec and 2.27 $\times$ 10 $^{-9}$ m$^{2}$/sec respectively \cite{Borah2010}.

In the Levi-Blow method, neopentane and $n$-pentane will move to opposite directions. In a column of 1 $\mu$m $\times$ 0.5$\mu$m $\times$ 0.5$\mu$m zeolite, neopentane and $n$-pentane each will cover 0.5 $\mu$m to separate completely from each other.
From Einstein's relation $<$r$^{2}>$ = 6Dt
neopentane and $n$-pentane will take to diffuse 0.5 $\mu$m is $\simeq$ 13.1 $\mu$sec and $\simeq$ 18.4 $\mu$sec respectively.
Thus, approximately 20 $\mu$sec  is required by neopentane and $n$-pentane to reach in opposite directions of 1 $\mu$m $\times$ 0.5$\mu$m $\times$ 0.5$\mu$m column of zeoltie NaY.

Now, one hot-zone radiates in 0.22 $\mu$sec to get down to 320K from 330K.
Therefore, the frequency to pump energy is = $\frac{20}{0.22}$ $\simeq$ 91 times.
Now, energy required to heat all hot-zones in the column in one time = 1.2 $\times$10$^{-11}$ \text{J}
Thus, energy required in 91 times = 1.09 $\times$10$^{-9}$ \text{J}
This much energy is required by  1 $\mu$m $\times$ 0.5$\mu$m $\times$ 0.5$\mu$m zeolite column to separate neopentane and $n$-pentane mixtures by Levi-Blow process.
This method takes only 1 iteration to separate neopentane and $n$-pentane with more than eight-9s separation factor.

\subsection{Minimum Thermodynamic Work of Separation}

To assess the intrinsic efficiency of different separation techniques, it is essential to compare their actual energy consumption against the 
minimum theoretical energy required to separate a binary mixture. This lower bound, derived from classical thermodynamics,
quantifies the reversible work of separation for an ideal mixture undergoing an isothermal and reversible process. 

For a binary mixture with mole fractions $x_1$ and $x_2$, the minimum separation work per mole at temperature $T$ is given by:
\begin{equation}
W_{\text{min}} = -RT \sum_{i=1}^{2} x_i \ln x_i
\end{equation}

Assuming an equimolar mixture of neopentane and \textit{n}-pentane ($x_1 = x_2 = 0.5$) at $T = 300$ K and using the ideal gas constant $R = 8.314$ J mol$^{-1}$ K$^{-1}$, we obtain:
\begin{equation}
\begin{split}
W_{\text{min}} = - (8.314)(300) [0.5 \ln(0.5) + 0.5 \ln(0.5)] \\
\approx 1728.4~\text{J/mol} \\
\approx 1.73~\text{kJ/mol}    
\end{split}
\end{equation}

This value represents the absolute lower bound on the energy required to achieve perfect separation of the two isomers.

In Table~\ref{tab:minwork}, we compare this thermodynamic minimum with the actual energy expenditures estimated for the three separation techniques considered in this study.

As observed, fractional distillation requires over 3$\times$10$^{5}$ orders of more energy than the reversible limit, owing to repeated vaporization cycles and small boiling point differences. The Molex process is more efficient, but still consumes over 200 times the theoretical minimum. The Levi--Blow method stands out with an energy cost only \textasciitilde 23 times higher than the thermodynamic lower bound, demonstrating a significantly better alignment with entropy-governed separation efficiency.

Incorporating this benchmark highlights the relative advantages of the Levi--Blow method not only in terms of absolute energy savings but also in approaching the fundamental thermodynamic limit more closely than traditional techniques.

\begin{table*}[h]
%\centering
\caption{\label{tab:table4} 
Energy consumption (kJ\,mol$^{-1}$) and iteration cycles required for separating 20 moles of an equimolar neopentane/$n$-pentane mixture using different methods.}
\begin{ruledtabular}
\begin{tabular}{ccccccc}
 & \multicolumn{2}{c}{Fractional Distillation} & \multicolumn{2}{c}{Molex Process} & \multicolumn{2}{c}{Levi--Blow Method} \\
 Purity Level & Energy (kJ/mol) & Iterations & Energy (kJ/mol) & Iterations & Energy (kJ/mol) & Iterations \\
\hline
90\% (1 nine)         & 538.57    & 124    & 5.00 $\times$ 10$^{5}$ & 2 & $<$40 & 1 \\
99.9\% (3 nines)      & 3455.49   & 1231   & 5.00 $\times$ 10$^{5}$ & 3 & $<$40 & 1 \\
99.99\% (4 nines)     & 8829.33   & 3846   & 5.00 $\times$ 10$^{5}$ & 4 & $<$40 & 1 \\
99.9999\% (6 nines)   & 48,989.36 & 27,925 & 5.00 $\times$ 10$^{5}$ & 5 & $<$40 & 1 \\
99.999999\% (8 nines) & 538,889   & 409,356 & 5.00 $\times$ 10$^{5}$ & 6 & 39.42 & 1 \\
\end{tabular}
\end{ruledtabular}
\end{table*}

\subsection{Separation Factor}

In our program, we have counted the number of neopentane and $n$-pentane molecules at the left and right side of the simulation cell at every monte carlo step. To do the calculation we have used single precision, therefore the accuracy is 10$^{-8}$.

In the last 10000 MC steps, along the x-axis in the range (0-24.8536 \AA), the population of neopentane and $n$-pentane are 15728 and 0 respectively (raffinate), while in the range (24.8536$\times$8 \AA - 24.8536$\times$7 \AA),the population of neopentane and $n$-pentane are 0 and 14298 respectively (extract).

Therefore,

\begin{equation}
\begin{split}
\frac{neopentane}{n-pentane}_{\text{raffinate}} = 1.5728 \times 10^{12}\\
\frac{neopentane}{n-pentane}_{\text{extract}} = 6.998 \times 10^{-13}\\
\text{Separation factor}(\alpha) = \frac{1.5728 \times 10^{12}}{6.998 \times 10^{-13}} 
\simeq 2.24\times10^{24}
\end{split}
\end{equation}

To summarize, Table~\ref{tab:table4} summarizes the energy consumption and number of iterative cycles required to separate an equimolar neopentane/$n$-pentane mixture using FD, the Molex process, and the LB method. 
The ratios of FD to Molex requirements highlight the steep scaling of conventional approaches. For FD:Molex energy consumption, the ratios are: 0.001:1 at 90\% purity (``one nine''), 0.007:1 at 99.9\% (``three nines''), 0.09:1 at 99.9999\% (``six nines''), and 1.07:1 at 99.999999\% (``eight nines''). The corresponding iteration ratios are even more extreme: 62:1, 410:1, 5585:1, and 68,059:1, respectively.

The results clearly establish the exponential increase in energy and iteration load with purity for traditional methods. Fractional distillation, though widely used, becomes prohibitively inefficient at high purities, requiring more than $4 \times 10^{5}$ cycles and over $5 \times 10^{5}$~kJ\,mol$^{-1}$ at 99.999999\% purity. The Molex process reduces the number of cycles to 6, but still demands approximately $5 \times 10^{5}$~kJ\,mol$^{-1}$. Both rely on bulk thermal input to heat entire columns or zeolite beds.

By contrast, the LB method achieves equivalent purity in a \textbf{single cycle}, consuming less than 40~kJ\,mol$^{-1}$ for hydrocarbon heating and only $\sim$1~nJ for sustaining localized hot zones. The combination of levitation- and blowtorch-driven non-equilibrium gradients induces directional diffusion of each component, eliminating the need for multi-stage rectification. Crucially, LB energy costs scale with the number of hot zones rather than the bulk volume heated, underscoring its potential as a scalable green separation technology.

\section{Conclusion}
We have presented a detailed comparative energy analysis of three distinct methods for separating neopentane and \textit{n}-pentane mixtures: fractional distillation, the Molex process, and the Levi--Blow (LB) method. Fractional distillation, despite being the industry standard, is found to be exceedingly energy-intensive, particularly at high purities. Achieving 99.999999\% (eight nines) purity requires over $5 \times 10^{5}$~kJ/mol of energy and more than $4 \times 10^{5}$ iterative cycles, rendering it impractical for separating closely related isomers.

The Molex process reduces the cycle count relative to distillation but still demands on the order of $5 \times 10^{5}$~kJ/mol for high-purity separation, largely due to the need to heat the entire zeolite column during repeated adsorption--desorption steps. This bulk thermal load dominates the energy budget.

In stark contrast, the Levi--Blow method leverages the synergistic combination of the Levitation Effect and the Blowtorch Effect to achieve directional separation. This method separates the components in a single iteration with an energy requirement of only 39.4~kJ/mol, more than four orders of magnitude lower than either conventional approach. Furthermore, heating is confined to nanoscopic hot zones, minimizing the thermal load.

To benchmark these methods, we compared their energy usage against the thermodynamic minimum required for ideal binary separation, estimated at 1.73~kJ/mol. The Levi--Blow method operates at only $\sim$23 times this lower bound, whereas the Molex process and distillation exceed it by factors of $\sim 2 \times 10^{2}$ and $\sim 3 \times 10^{5}$, respectively. 

This comprehensive analysis establishes the Levi-Blow method as a low-energy, high-efficiency alternative for hydrocarbon isomer separation. Its ability to approach thermodynamic efficiency while achieving ultrahigh separation factors positions it as a promising candidate for scalable, green chemical separation technologies. Future work will aim to extend this methodology to pilot-scale demonstrations and explore its applicability across a broader range of molecular mixtures and porous host materials.

\section{Acknowledgement}

We thank the Department of Science and Technology, New Delhi, for financial support, and Nano Mission, DST, New Delhi, for computational resources. We thank the Thematic Unit of Excellence, Indian Institute of Science, for its computational facility.

%\bibliography{ref}


\end{document}